\documentclass{ws-procs9x6}

\begin{document}

\title{
\vskip -25mm
{\hfill \small DESY 02-169 \\
\hfill \small EFI 02-55} \\
\vskip 15mm
Production of Missing \lowercase{$c\bar{c}$ and $b\bar{b}$}
States\footnote{\uppercase{T}his work was supported 
in part by the \uppercase{U}nited
\uppercase{S}tates \uppercase{D}epartment of \uppercase{E}nergy 
through \uppercase{G}rant \uppercase{N}o.\ \uppercase{DE FG02 
90ER40560}
and the \uppercase{N}atural \uppercase{S}ciences and \uppercase{E}ngineering 
\uppercase{R}esearch \uppercase{C}ouncil of \uppercase{C}anada.}}

\author{Stephen Godfrey$^{\lowercase{a,b}}$
%\footnote{E-mail: godfrey@physics.carleton.ca}
and Jonathan L. Rosner$^{\lowercase{c}}$
%\footnote{E-mail: rosner@hep.uchicago.edu}
}

\address{$^a$Department of Physics,
Carleton University, Ottawa K1S 5B6 CANADA \\
$^b$DESY, Deutsches Elektronen-Synchrotron, D22603 Hamburg, GERMANY \\ 
%}
%\author{Jonathan L. Rosner\footnote{E-mail: rosner@hep.uchicago.edu}}
%\address{
$^c$Enrico Fermi Institute and Department of Physics \\
University of Chicago, 5640 S. Ellis Avenue, Chicago, IL 60637}

\maketitle

\abstracts{
The heavy quarkonium $c\bar{c}$ and $b\bar{b}$ resonances have a rich
spectroscopy with numerous narrow $S$, $P$, and $D$-wave levels below
the production threshold of open charm and beauty mesons.  The mass
predictions for these states are an important test of QCD
calculations. % Many heavy quarkonium resonances remain undiscovered.
We review some recent work describing the production of
missing $\eta_b(nS)$, $1^3D_J(b\bar{b})$ and $1^1P_1(c\bar{c})$ and
$1^1P_1(b\bar{b})$ states. }

%\section{Introduction}

The recent discovery of a D-wave 
$b\bar{b}$ state by the CLEO collaboration \cite{cleo02} 
and the discovery of the $\eta_c'$ in $B\to \eta_c' K$
by the BELLE collaboration \cite{belle02} have brought
heavy quarkonium physics to the forefront.
The CLEO collaboration continues to take data at the $\Upsilon(3S)$ and 
$\Upsilon(2S)$, and the B-factories'
observation of charmonium states in $B$ decay has led to the hope that some 
of the missing $c\bar{c}$ states may be observed in $B$ decay.
At the same time, advances in lattice QCD calculations of the heavy 
quarkonium spectra are leading to quantitative predictions which need 
to be tested against experiment \cite{lattice}.
With these developments, we expect a 
much better understanding of heavy quarkonium.
We summarize some recent work describing the 
production of missing $1^3D_J(b\bar{b})$, $\eta_b(nS)$ and 
$1^1P_1(c\bar{c})$ and $1^1P_1(b\bar{b})$ states\cite{gr-dw,gr-etab,gr-hb}.

Although $^3D_1$ states can be produced in $e^+e^-$ collisions this approach
requires substantial statistics owing to the small coupling of the
$^3D_1(b\bar{b})$ to the photon and the effects of line smearing.  
Another approach is via the
electromagnetic cascades $\Upsilon(3S) \to \gamma \chi_b' \to \gamma 
\gamma \; ^3D_J$.  Using quark model estimates of both E1 dipole 
transitions and decays to final state hadrons \cite{KR} we 
estimate the number of events for $4\gamma$ cascades that proceed 
via various intermediate states \cite{gr-dw} with the dominant 
cascades listed in 
Table 1.  We expect a total of $\sim 38$ events per $10^6$ 
$\Upsilon(3S)$'s to be produced via $^3D_J$ states.  
A substantial background will be $4\gamma$ cascades 
proceeding via the $2^3S_1$ state 
which will also produce $\sim 38$ events per $10^6$ $\Upsilon(3S)$'s.  These 
events can be separated from those with $1^3D_J$'s by taking 
advantage of the different intermediate photon
energies.  The  CLEO collaboration
successfully employed this strategy for the first observation 
of a triplet $\Upsilon(1D)$ state \cite{cleo02}.

\begin{table}[ph]
\tbl{Predicted numbers of some $4 \gamma \; e^+ e^-$ cascade 
events 
per $10^6~\Upsilon(3S)$ decays.  From Ref. [4]. %\cite{gr-dw}.
\vspace*{1pt}
}
{\footnotesize
\begin{tabular}{|l | l|} 
\hline
{} &{}\\[-1.5ex]
Cascade & Events \\[1ex]
\hline
{} &{}\\[-1.5ex]
$3^3S_1 \to 2^3P_2 \to 1^3D_3 \to 1^3P_2 \to 1^3S_1$ & 7.8 \\[1ex]
$3^3S_1 \to 2^3P_2 \to 1^3D_2 \to 1^3P_1 \to 1^3S_1$ & 2.7 \\[1ex]
$3^3S_1 \to 2^3P_1 \to 1^3D_2 \to 1^3P_1 \to 1^3S_1$ & 20 \\[1ex]
$3^3S_1 \to 2^3P_1 \to 1^3D_1 \to 1^3P_1 \to 1^3S_1$ & 3.3 \\[1ex]
\hline
\end{tabular}  
\label{tab1} 
}
\vspace*{-13pt}
\end{table}

%\section{$\eta_b(nS)$ States}

The $\eta_b(nS)$'s can be produced via the magnetic dipole (M1) transitions
$\Upsilon(nS)\to \eta_b(n' S) +\gamma$ \cite{gr-etab}.  The available phase 
space for allowed transitions (principal quantum number 
is unchanged) is small resulting in a small partial width.
The hindered transitions (principal quantum 
number changes) have large available phase space but in the nonrelativistic 
limit the overlap integral is zero owing to the 
orthogonality of the initial and final wavefunctions.  However, 
relativistic corrections from the hyperfine interaction lead to
differences in the $^3S_1$ and $^1S_0$ wavefunctions  resulting in a 
non-zero overlap.  The branching ratios for 
$\Upsilon(nS)\to \eta_b(n'S) \gamma $ are given in Table 2.  
With these BR's we expect $\eta_b$'s to be produced at a substantial 
rate.  Another possible production process for $\eta_b$'s is 
$\Upsilon(3S)\to h_b (^1P_1) \pi \pi \to \eta_b + \gamma +\pi 
\pi$\cite{ky81,voloshin}.  
The BR for  $\Upsilon(3S)\to h_b (^1P_1) \pi \pi$ is expected 
to be 0.1-1~\% while the BR for  $h_b (^1P_1) \to \eta_b + \gamma$ is 
estimated to be $\sim 50\%$.

\begin{table}[tph]
\tbl{
Branching ratios of M1
transitions between $n^3S_1$ and $n'^1S_0$ $b \bar b$ levels taking into
account relativistic corrections. From Ref. [5]
%\cite{gr-etab} 
using wavefunctions from Ref. [8].
%\cite{gi85}.
\vspace*{1pt}}
{\footnotesize
\begin{tabular}{|l c c|} 
\hline
{} &{} &{}\\[-1.5ex]
 & Transition \qquad & $B(10^{-4})$ \\[1ex]
\hline
{} &{} &{}\\[-1.5ex]
$\Upsilon (3S)$ 		& $\to 3^1S_0$ & 0.10 \\[1ex]
				& $\to 2^1S_0$ & 4.7 \\[1ex]
				& $\to 1^1S_0$ & 25 \\[1ex]
$\Upsilon (2S)$ 		& $\to 2^1S_0$ & 0.21 \\[1ex]
				& $\to 1^1S_0$ & 13 \\[1ex]
$\Upsilon (1S)$ 		& $\to 1^1S_0$ & 2.2 \\[1ex]
\hline
\end{tabular}\label{tab2} }
\vspace*{-13pt}
\end{table}

%\section{Production of $^1P_1$ States}

To produce the $^1P_1$ $c\bar{c}$ and $b\bar{b}$ states
we start by considering the electromagnetic cascade\cite{gr-hb};
\begin{equation}
\Upsilon(3S)\stackrel{M1}{\to} \eta_b(2S) +\gamma \stackrel{E1}{\to} 
h_b(1P) +\gamma\gamma \stackrel{E1}{\to} \eta_b(1S) 
+\gamma\gamma\gamma
\end{equation}
Quark model estimates of the branching fractions give\cite{gr-hb}
$\Gamma[\eta_b(2S)\to h_b(1P)+\gamma] =2.3$~keV
and $\Gamma [\eta_b(2^1S_0) \to gg]= 4.1 \pm 0.7$~MeV.
Combining the resulting  branching ratios for 
${B}[\Upsilon(3S)\to \eta_b' + \gamma]$ and ${B}[\eta_b'\to h_b 
+\gamma]$ gives 
${B}[\Upsilon(3S)\to \eta_b' \gamma \to h_b 
\gamma\gamma]=2.6\times 10^{-7}$ resulting in only 0.3 
events per $10^6 \; \Upsilon(3S)$'s.  Similarly,
${B}[\psi(2S)\to \eta_c' \gamma \to h_c \gamma\gamma]=10^{-6}$ or 1 
event per $10^6 \; \psi'$'s.  
A more promising possibility utilizes
the hadronic decay $\Upsilon(3S)\to 
\pi 1^1P_1$ which is estimated to have a BR of around 0.1\%\cite{voloshin}:
\begin{equation}
\Upsilon(3S)\to h_b(1P) +\pi \stackrel{E1}{\to} \eta_b(1S) +\gamma +\pi
\end{equation}
The  radiative and hadronic widths of the $^1P_1$ states 
given by Ref. [6] %\cite{gr-hb} are
$\Gamma[h_b(1P) \to\eta_b(1S)+\gamma]=37$~keV and
$\Gamma[h_b(1P) \to ggg] = 50.8$~keV.
Combining the branching ratios in this decay chain yields
${B}[\Upsilon(3S)\to  h_b + \pi^0 \to \eta_b + \gamma \pi^0]= 
4 \times 10^{-4}$
resulting in 400 events per $10^6$ $\Upsilon(3S)$'s.  Similarly
${B}[\psi(2S)\to  h_c + \pi^0 \to \eta_c +\gamma \pi^0]= 3.8 
\times 10^{-4}$.
In both cases it appears that the $1^1P_1$ should be produced in 
sufficient numbers to be observed.  
The recent Belle observation\cite{belle02} 
of the $\eta_c(2S)$ in $B\to \eta_c(2S) K$
also offers the possibility that the $h_c$ can also be observed in $B$ decay.

\end{document}